\documentclass[11pt]{article}
\usepackage{amsmath, amsthm, amssymb, graphicx}
\usepackage{psfrag}
\usepackage{cite}
\usepackage{epsfig}
\usepackage{mathrsfs}

\newcommand{\p}{\partial}
\newcommand{\s}{\sigma}

\renewcommand{\u}{{\bf u}}
\newcommand{\f}{c}

\newcommand{\al}{\alpha}
\newcommand{\x}{{\bf x}}
\newcommand{\D}{\nabla}

\newcommand{\tf}{\tilde{\f}}
\newcommand{\tu}{\tilde{{\bf u}}}
\newcommand{\g}{\gamma}
\newcommand{\Mi}{M_{\text{iso}}}
\newcommand{\Mu}{M_{\text{uni}}}
\newcommand{\fepo}{Li$_x$FePO$_4$}
\newcommand{\F}{\mathscr{F}}

\begin{document}

\title{{\huge Phase Separation with Anisotropic Coherency Strain }}
\author{Liam G. Stanton \\
{\it  Center for Applied Scientific Computing }\\ 
{\it Lawrence Livermore National Laboratory, 
Livermore, CA 94550 USA } \\
\ \\
Martin Z. Bazant\footnote{ bazant@mit.edu } \\
{\it Departments of Chemical Engineering and Mathematics} \\
{\it Massachusetts Institute of Technology, Cambridge, MA 02139 USA }
}

\date{\today\footnote{ This paper was originally written in December 2008. Only minor changes have been made, mainly to discuss more recent developments on \fepo. } }
\maketitle{}

\begin{abstract}
We consider the effects of anisotropic coherency strain (due to lattice mismatch) on phase separation in intercalation materials, motivated by the high-rate Li-ion battery material \fepo. Using a phase-field model coupled to elastic stresses, we analyze spinodal decomposition (linear instability of the homogeneous state) as well as nonlinear evolution of the phase pattern at constant mean filling. We consider fully anisotropic coherency strain and focus on the case of simultaneous expansion and contraction along different crystal axes, as in the case of \fepo, which leads to tilted, striped phase boundaries in equilibrium. 
\end{abstract}

\section{Introduction}

Intercalation, i.e. the reversible insertion of  molecules in a host solid, is a general phenomenon with many applications in chemical physics. Of particular interest today is the intercalation of lithium ions in the active solid particles of Li-ion battery composite electrodes.  Li-ion batteries have become essential for portable 
electronic devices such as cell phones and power tools and are advancing toward possible use in electric vehicles or renewable energy storage, due to higher energy density and power density than earlier battery chemistries \cite{megahed94, wakihara98, scrosati00}.  In particular, \fepo~ has gained considerable 
attention as a leading cathode material candidate because of its high 
power density, fast charging times, long cycle life, and favorable safety characteristics \cite{padhi97, 
tarascon01, kang09}.

Modeling ion intercalation dynamics in crystalline particles is a complex problem at the intersection 
of  physics, chemistry, materials science, and applied mathematics  \cite{newman91, bard01, 
rubinstein90}. A standard assumption in Li-ion battery models is that diffusion is linear and isotropic within the solid  \cite{newman91, 
doyle93}, but it is becoming recognized that this is often not the case, due to the anisotropic crystalline nature of most solid hosts. In particular, experiments and {\it ab initio} computer simulations have revealed 
that Li diffusion in \fepo~ is highly {\it anisotropic} and exhibits a strong tendency for  
phase-separation between lithiated (LiFePO$_4$) and delithiated (FePO$_4$)
states \cite{laffont06, amin07,Ramana2009}.  Additionally, this FePO$_4$/LiFePO$_4$ 
phase boundary tends to reach the active (010) surface and align itself along \{101\} or \{100\} crystal planes  \cite{laffont06,chen06,Ramana2009}, which suggests an intimate connection between phase separation dynamics and the elastic properties of the crystal. In contrast, motivated by the first experimental paper~\cite{padhi97}, existing battery engineering models for \fepo assume
a spherical ``shrinking core" of one phase being replaced by a shell of the other~\cite{srinivasan04,dargaville2010} and thus do not model phase separation dynamics, anisotropic diffusion or coherency strain. 

Over the past five years, our group has developed a general phase-field model for intercalation kinetics in nanoparticles, which is helping to unravel the complex dynamical behavior of \fepo and interpret experimental data from microscopic first principles~~\cite{Bazant2012,singh08,burch2008,burch2009,Bai2011,Cogswell2012,Bazant_notes}. A crucial contribution of this work has been to consistently  formulate the kinetics of electrochemical charge-transfer reactions at the surface of the active particles with the non-equilibrium thermodynamics of bulk ion transport and phase separation via a modified Butler-Volmer theory for concentrated solutions and solids~\cite{Bazant2012,Bazant_notes}. This has led to novel predictions for intercalation dynamics in nano particles, such as size-dependent miscibility and spinodal gaps~\cite{burch2009,Wagemaker2011}, ion intercalation waves filling the crystal layer by layer (rather than  shrinking core) at low currents~\cite{singh08,burch2008,Tang2011}, and quasi-solution behavior and suppressed phase separation at high currents~\cite{Bai2011,Cogswell2012}. 

In this work, we focus on understanding the general effects of coherency strain on bulk phase separation, while neglecting the coupling to electrochemical reactions.  (For a complete model of \fepo nanoparticles under applied current, see Ref.~\cite{Cogswell2012}.) Classical phase-field models in materials science have included effects of elastic coherency strain for isotropic solids~\cite{cahn61,cahn62b,larche85,gurtin96}, but here we analyze phase separation with anisotropic coherency strain, focusing on the novel case where both contraction and expansion occur along different crystal planes, as in \fepo.  Simplified elastic models of \fepo  have led to the conclusion that phase separation occurs along the \{100\} crystal planes \cite{meethong07a,VanderVen2009,Tang2011}, as in some experiments~\cite{chen06}, even though \{101\} phase boundaries are also sometimes observed \cite{laffont06,Ramana2009}. Here we show that such tilted phase boundaries result from simultaneous contraction and expansion in different directions. 
To address dynamical behavior, we perform a linear stability analysis on the model to predict parameter regimes for
spinodal decomposition of the system, which provides analytical predictions for initial
instabilities in the concentration and deformation fields.  As the nonuniform states are
highly nonlinear, numerical simulations are necessary to examine the long-term 
behavior of phase-separated solutions.
 
\section{Model Formulation}
To develop a phase-field model of this system,  
we begin by defining an order parameter $\f$ to be the
dimensionless, normalized concentration of Li, where 
$\f=1$ corresponds to an entirely lithiated phase
(LiFePO$_4$), and $\f=0$ to an entirely delithiated one (FePO$_4$).
The total free energy of the system described by the domain $\Omega$
can be expressed as a functional of the local Li concentration in
the Cahn-Hilliard formulation
\begin{equation}\label{Ftot}
    \F[\f] = \int_\Omega\left\{f(\f) + \frac{\rho}{2}\nabla\f \cdot
    ({\bf K}\nabla\f) + \frac{1}{2}{\bf E}:{\bf T}\right\}\text{d}{\bf x},
    \quad \x\in\Omega,
\end{equation}
where $f(\f)$ is the free energy density of a homogeneous state, 
$\rho$ is the number density of intercalation sites, and
the gradient term (with ${\bf K}$ being a symmetric, positive
definite tensor) is the next order term in an expansion about this
state which penalizes spatial phase modulation in the system
\cite{cahn58,Nauman2000}. For our model, we take this tensor to be constant. 
Li insertion into this material has been observed to
produce significant anisotropic strains which range from about -2\% to 5\% 
to maintain coherency between states \cite{meethong07a}, and hence 
we also take into account long-range elastic
contributions to the free energy as in previous phase-field models
\cite{larche85, gurtin96}. The elastic energy is
represented in the final term of the integrand in Equation (\ref{Ftot}) 
with the contraction between the tensors ${\bf E}$ and ${\bf
T}$, which are the strain and stress tensors respectively. 
To capture the effects of the interactions between lithiated and
delithiated states, we take the regular solution model for the
homogeneous energy density \cite{cahn58}
\begin{equation}\label{homog}
    f(\f) = a\rho\f(1-\f) + \rho k_BT\left[\f\log(\f) +
    (1-\f)\log(1-\f)\right],
\end{equation}
where $k_B$ is the Boltzmann constant, $T$ is the absolute temperature, and 
the regular solution interaction parameter $a$ is the average energy 
density of ion intercalation. The first 
term in Equation (\ref{homog}) is the enthalpic contribution which favors phase
separation, while the second term is the entropic contribution which
promotes phase mixing. The elements of the adjusted infinitesimal strain tensor
are given by the kinematic relationship
\begin{equation}\label{kinematic}
    {\bf E} = \frac{1}{2}\left(\D\u + \D\u^{\text{T}}\right) - \f{\bf M},
\end{equation}
where $\u$ is the deformation vector, and ${\bf M}$ is the lattice
mismatch tensor of the lithiated and unlithiated phases. This
additional mismatch term is the stress-free strain when the
density is $\f$. Noting that the intercalation of these particles takes the
delithiated orthorhombic crystal (FePO$_4$) to another orthorhombic state 
(LiFePO$_4$), insertion will hence only induce compressions and expansions 
of the surrounding material.  Consequently, the tensor ${\bf M}$ will be diagonal. 
Assuming the strains to be within the regime of linear elasticity, we can then
use Hooke's law to describe the stress in terms of the constitutive equation
\begin{equation}\label{constitutive}
    {\bf T} = {\bf C}:{\bf E},
\end{equation}
where ${\bf C}$ is the rank-4 elasticity tensor \cite{landau59}, which for orthorhombic 
crystals, can be described by nine distinct elements \cite{maxisch06}.
We now use this free energy to calculate the chemical potential as
\begin{equation}
    \mu \equiv \frac{1}{\rho}\frac{\delta\F}{\delta\f} =
    \frac{1}{\rho}\frac{\p}{\p\f}
    \left[f(\f)+\frac{1}{2}{\bf E}:{\bf T}\right] -
    \nabla\cdot ({\bf K}\nabla\f),
\end{equation}
where
\begin{align}
    \frac{\p f}{\p\f} = \,a\rho(1-2\f) + \rho k_BT\log\left(
    \frac{\f}{1-\f}\right),\quad
    \frac{1}{2}\frac{\p}{\p\f}\left({\bf E}:{\bf T}\right)
    = -{\bf M}:{\bf T}.
\end{align}
Note that as ${\bf M}$ is diagonal, only the compressional
components of the stress will affect the chemical potential. The mass
flux is then expressed as
\begin{equation}
    {\bf j} = -\rho\f{\bf B}\nabla\mu,
\end{equation}
where ${\bf B}$ is the mobility tensor. Strictly speaking, the mobility must depend on concentration, e.g. as ${\bf B} \sim (1-\f)$ in the classical Cahn-Hilliard regular solution model~\cite{Nauman2000}, but below we will   make simpler assumptions that have relatively little effect on the dynamics of coherency phase separation.

Finally, the evolution of the concentration field is calculated with
a conservation law within the cathode
\begin{equation}
    \rho\frac{\p\f}{\p t} + \nabla\cdot{\bf j} = 0,
\end{equation}
and at any given time, a quasi-steady approximation of mechanical
equilibrium is given by
\begin{equation} \label{stress}
    \D\cdot{\bf T} = {\bf 0}.
\end{equation}
Using Equations (\ref{kinematic}), (\ref{constitutive}) and (\ref{stress}), we can 
write this elastostatic condition in terms of the deformation field components as 
the system

\begin{subequations}\label{NCeqns}
\begin{align}
	&\frac{\p}{\p x}\sum_{k=1}^3C_{1k}\frac{\p u_k}{\p x_k} + C_{66}\frac{\p}{\p y}
	\left(\frac{\p u_1}{\p y} + \frac{\p u_2}{\p x}\right) + C_{55}\frac{\p}{\p z}
	\left(\frac{\p u_3}{\p x} + \frac{\p u_1}{\p z}\right) = \sum_{k=1}^3C_{1k}M_{kk}
	\frac{\p\f}{\p x},\\
	&\frac{\p}{\p y}\sum_{k=1}^3C_{2k}\frac{\p u_k}{\p x_k} + C_{44}\frac{\p}{\p z}
	\left(\frac{\p u_2}{\p z} + \frac{\p u_3}{\p y}\right) + C_{66}\frac{\p}{\p x}
	\left(\frac{\p u_1}{\p y} + \frac{\p u_2}{\p x}\right) = \sum_{k=1}^3C_{2k}M_{kk}
	\frac{\p\f}{\p y},\\ \label{NCeqns3}
	&\frac{\p}{\p z}\sum_{k=1}^3C_{3k}\frac{\p u_k}{\p x_k} + C_{55}\frac{\p}{\p x}
	\left(\frac{\p u_3}{\p x} + \frac{\p u_1}{\p z}\right) + C_{44}\frac{\p}{\p y}
	\left(\frac{\p u_2}{\p z} + \frac{\p u_3}{\p y}\right) = \sum_{k=1}^3C_{3k}M_{kk}
	\frac{\p\f}{\p z},
\end{align}
\end{subequations}
where we are using the standard convention to reduce the elasticity tensor elements 
$C_{ijkl}$ to the matrix components $C_{ij}$ \cite{nye85}. This hence gives 
the elastic energy contribution to the chemical potential
\begin{align}
    \frac{1}{2}\frac{\p}{\p\f}\left({\bf E}:{\bf T}\right)
    = -{\bf M}:{\bf T} = \sum_{i=1}^3\left[\sum_{j=1}^3C_{ij}M_{jj}
    \left(M_{ii}\f - \frac{\p u_i}{\p x_i}\right)\right].
\end{align}
With the bulk equations established for Li migration within the
cathode, boundary conditions must be formulated at the
electrode-electrolyte interfaces to close the model. Mass balance at 
the interface must satisfy
\begin{equation}
    {\bf n}\cdot{\bf j} + R(\f,\mu) = 0,
\end{equation}
where ${\bf n}$ is the outward normal to the surface, and $R(\f,\mu)$ is the
potential-dependent interfacial reaction rate~\cite{Bazant2012,burch2009,Bai2011,Cogswell2012,Bazant_notes}. We additionally assume the so-called 
variational boundary condition
\begin{equation}
    {\bf n}\cdot({\bf K}\D\f) = 0,
\end{equation}
which is reasonable for our system in which concentration gradients
are observed to be perpendicular to the particle surface \cite{burch09}. 
Finally, we assume a uniform pressure field of strength $P$ from the
surrounding matrix, which gives the interfacial stress balance
\begin{equation}
    {\bf n}\cdot({\bf T} + P{\bf I}) = {\bf 0}.
\end{equation}

It should be noted that we state the boundary conditions above just to show the form of the complete model, but we will not use them in our calculations below. 
As the particular focus of this work is a feature
of the bulk dynamics, we take the infinite domain $\Omega = \mathbb{R}^3$, and 
hence only boundedness in the far-field of the concentration and deformation fields is 
necessary for the purposes of our analysis.  Detailed studies of current dependence and finite size effects for ion intercalation in nanoparticles can be found in \cite{Bai2011,Cogswell2012,burch2009,Wagemaker2011,Tang2011,meethong07b}.

\section{Isotropic Coherency Strain}
To begin to understand the leading order effects of lithiation induced strains
in \fepo, we start by taking the host material to be an isotropic body. Under this
assumption, the coefficients of the elasticity tensor reduce to
\begin{align}
	&C_{11} = C_{22} = C_{33} = 2\left(\frac{1-\nu}{1-2\nu}\right)G,\quad
	C_{44} = C_{55} = C_{66} = G,\\
	&C_{12} = C_{21} = C_{23} = C_{32} = C_{13} = C_{31} = \left(
	\frac{2\nu}{1-2\nu}\right)G,
\end{align}
where $G$ is the shear modulus, and $\nu$ is the Poisson ratio. We further
take the coherency strains to be isotropic, which simplifies the mismatch tensor to
the form ${\bf M} = \Mi{\bf I}$, where $\Mi$ is a scalar. We can then write the elastostatic 
condition (\ref{stress}) in terms of the deformation vector as
\begin{align}\label{eqnU}
    (1-2\nu)\D^2\u + \D(\D\cdot\u)  = 2\Mi(1+\nu)\D\f.
\end{align}
Note that this also simplifies the elastic contribution to the
chemical potential to
\begin{equation}
    \frac{1}{2}\frac{\p}{\p\f}\left({\bf E}:{\bf T}\right)
    = -\Mi\text{tr}\{{\bf T}\} = 2\Mi G\left( \frac{1+\nu}{1-2\nu}
    \right)\left[3\Mi\f - \D\cdot\u\right].
\end{equation}
If we take the curl of Equation (\ref{eqnU}), we obtain $\D^2(\D\times\u) = 0$.  
From potential theory, we know that $\u$ must then be curl free and can
be written in terms of the gradient of a scalar potential: $\u =
\D\psi$.  This gives the equation
\begin{equation}
    \D\left[\D^2\psi  - \Mi\left(\frac{1+\nu}{1-\nu}\right)\f\right] = {\bf 0}.
\end{equation}
This can then be integrated with the constant of integration taken
as zero without loss of generality. We hence have the dilatation, 
$\D\cdot{\u}$, as
\begin{equation}
    \D\cdot{\u} = \D^2\psi = \Mi\left(\frac{1+\nu}{1-\nu}\right)\f
\end{equation}
and can use this to rescale the enthalpic contribution in the 
chemical potential as
\begin{align}
    \mu = a - 2b\f + k_BT\log\left(\frac{\f}{1-\f}\right) -
    \nabla\cdot ({\bf K}\nabla\f),\quad b = a - 2\Mi^2
    \left(\frac{1+\nu}{1-\nu}\right)\frac{G}{\rho}.
\end{align}
Note that this result was first obtained and analyzed for phase-field models
in \cite{cahn61, cahn62b}. 

\begin{figure} \centering
  \resizebox{4in}{!}{\includegraphics[width=4.5in]{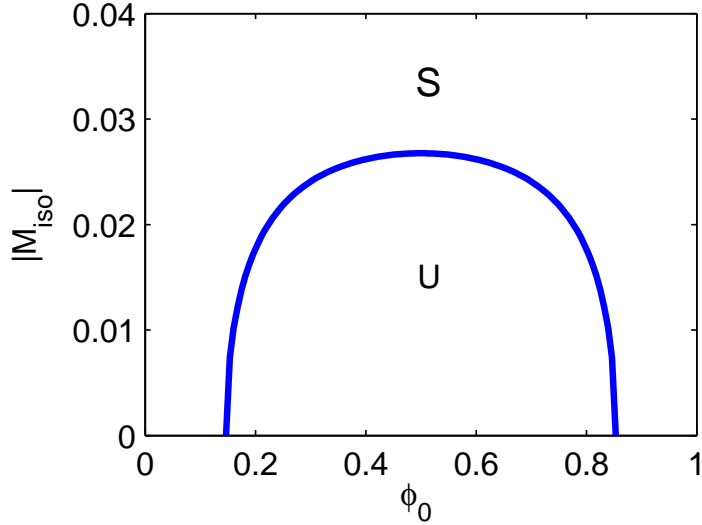}}
  \caption[Spinodal region in the parameter space]
  {Spinodal region in terms of the isotropic lattice mismatch $\Mi$ and 
    homogeneous concentration $\f_0$. The stable and unstable regions are 
    labeled within the figure.
    The parameters used are $a = 1.6 \times 10^{-20}\,
    \text{J}$ \cite{dodd06}, $\rho = 14\,\text{nm}^{-3}$, $G = 5 \times 10^{10} 
    \,\text{Pa}$, $\nu = 0.22$ \cite{maxisch06}, and $k_BT = 4 \times 
    10^{-21}\,\text{J}$ at room temperature.}
  \label{spin}
\end{figure}

We can next obtain a spinodal region in the parameter space, which is the region 
in which homogeneous concentrations are {\it linearly} unstable. 
The homogeneous state of the system can be calculated
by taking $\f = \f_0$ to be constant, and in perturbing this state with the
infinitesimal quantity $\tilde{\f} = \f - \f_0$, the linearized
evolution equation for the concentration perturbation will take the
form
\begin{align}
    \frac{\p\tf}{\p t} = \f_0\D\cdot\left\{{\bf B}\D\left[
      \left(-2b + \frac{k_BT}{\f_0(1-\f_0)}\right)\tf - 
      \nabla\cdot({\bf K}\nabla\tf)\right]\right\}.
\end{align}
By assuming the normal mode of the perturbation as $\tf =
\hat{\f}\exp(\sigma t + i{\bf q}\cdot\x)$, where $\sigma$ is growth
rate of a given wave vector ${\bf q}$, this gives the dispersion relation
\begin{equation}
    \sigma({\bf q}) = -\f_0{\bf q}\cdot({\bf B}{\bf q})\left[-2b + 
    \frac{k_BT}{\f_0(1-\f_0)} + {\bf q}\cdot({\bf K}{\bf q})\right].
\end{equation}
The spinodal can now be calculated for parameters in which there
exists a positive growth rate, and hence we have the region defined
by
\begin{equation}
    \Mi^2 < \frac{\rho}{4G}\left(\frac{1-\nu}{1+\nu}\right)\left[2a -  
      \frac{k_BT}{\f_0(1-\f_0)}\right].
\end{equation}
It is important to note that this stability criterion is independent of the 
sign of $\Mi$, and hence both crystal expansion and contraction will have
the same effect. Additionally, the increase of elastic strains narrows 
the stability region in parameter space and can even annihilate the spinodal 
for a given interaction parameter $a$ as shown in Figure (\ref{spin}). The
parameters obtained from experiment are listed in the caption. This latter point 
is somewhat troubling as not only does \fepo~exhibit both high strains and 
phase-separation upon intercalation, but the phase-separation in this material 
has been observed to occur along the direction of {\it greatest} strain 
\cite{chen06, rousse03}. This apparent contradiction is resolved in the following 
section.

\section{Anisotropic Coherency Strain}\label{anisomm}
\begin{figure} \centering
  \resizebox{4in}{!}{\includegraphics[width=4.5in]{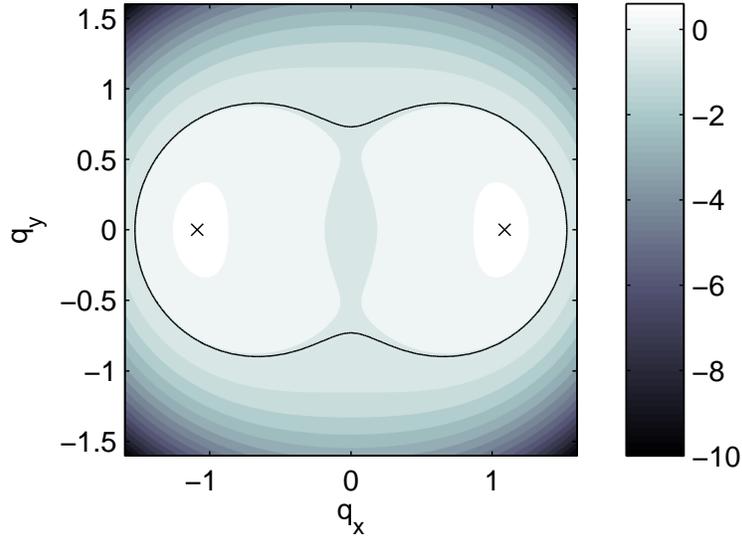}}
  \caption[Dispersion relation with anisotropic mismatch strains]
  {Dispersion relation with anisotropic mismatch strains. The
  curve indicates neutral stability ({\it i.e.~}$\s=0$), while the $\times$'s
  mark the largest growth rates. The lattice mismatch parameters 
  used are $M_{11} = 0.052$, $M_{22} = 0.036$ and $M_{33} = -0.019$ 
  \cite{maxisch06}. }
  \label{disp5}
\end{figure}

To account for the alignment of the FePO$_4$/LiFePO$_4$ phase boundary 
with the $(b,c)$ plane of the host crystal, we must include anisotropic properties of the
system into our model (note that we are using the standard notation for the axis 
labels of \fepo~as seen in \cite{chen06}).  In particular,
a more appropriate approximation for the mismatch tensor corresponding to this system
would be diagonal with distinct values. For \fepo, these diagonal 
elements are roughly: $M_{11} = 0.052$, $M_{22} = 0.036$ and $M_{33} = -0.019$ 
\cite{chen06}.  To capture the effects of anisotropic coherency strains, we will
simplify the model by reducing the remaining tensors to scalars as ${\bf B} = 
B{\bf I}$ and ${\bf K} = K{\bf I}$, as they merely act as spatial scalings. 
We can then easily nondimensionalize the spatial and temporal coordinates 
with the following relations
\begin{equation}
  \x \rightarrow \lambda \x,\quad t \rightarrow \frac{\lambda^2}{\mathcal{D}}
  t, \quad \lambda = \sqrt{\frac{K}{k_BT}}.
\end{equation}
where we have used the Einstein mobility relation to introduce the diffusivity 
$\mathcal{D} = k_BTB$.  The characteristic length-scale, $\lambda$, represents the 
typical width of the 
FePO$_4$/LiFePO$_4$ interface which is on the order of nanometers 
\cite{chen06}.  We hence obtain the concentration evolution
\begin{align}
  &\frac{\p\f}{\p t} = \D\cdot\left\{\f\D\left[-2\al\f + \log\left(\frac{\f}
    {1-\f}\right) - \D^2\f - \frac{1}{\rho k_BT}\sum_{i=1}^3
  \sum_{j=1}^3M_{ii}C_{ij}\frac{\p u_j}{\p x_j}\right]\right\},\\ 
  &\text{where}\quad \al = \frac{a}{k_BT} - \frac{1}{2\rho k_BT}\sum_{i=1}^3
  \sum_{j=1}^3M_{ii}C_{ij}M_{jj},
\end{align}
which is coupled with the equilibrium conditions (\ref{NCeqns}).
To further further illuminate the effects of coherency strain anisotropy, 
we again take the host material to be an isotropic body.  While this is not
quite physical, simulations have been performed for both isotropic and 
anisotropic bodies using the elastic moduli found in \cite{maxisch06} to verify 
this approximation, and no qualitative differences have been observed. Furthermore, 
as also seen in \cite{maxisch06}, the anisotropy is much stronger in the lattice mismatch 
({e.g.}~a {\it contraction} along the $c$-axis is induced by insertion), and hence
these effects are amplified by the fact that while the energetic contributions from 
elasticity are $\mathcal{O}(C_{ij})$, they are indeed quadratic in the mismatch 
parameters \cite{cahn62a}. This system therefore simplifies to
\begin{align}\label{bulk1}
  &\frac{\p\f}{\p t} = \D\cdot\left\{\f\D\left[-2\al\f + \log\left(\frac{\f}
    {1-\f}\right) - \D^2\f - 2\g\left({\bf M}\D + \frac{\nu}{1-2\nu}\text{tr}
    \{{\bf M}\}\D \right)\cdot\u\right]\right\},\\ \label{bulk2}
  &(1-2\nu)\D^2\u + \D(\D\cdot\u) = 2\left[(1-2\nu){\bf M}\D + \nu\,\text{tr}
    \{{\bf M}\}\D
    \right]\f,\\
  &\text{where}\quad \al = \frac{a}{k_BT} - \g\left[\text{tr}\{{\bf M}^2\} 
    + \left(\frac{\nu}{1-2\nu}\right)\text{tr}\{{\bf M}\}^2\right],\quad
  \g = \frac{G}{\rho k_BT},
\end{align}
and tr\{$\circ$\} denotes the trace operator.
As before, we perturb our system about the basic state of constant 
concentration and zero deformations to obtain the linearized system
\begin{align}
  &\frac{\p\tf}{\p t} = \f_0\D^2\left[\left(-2\al + \frac{1}{\f_0(1-\f_0)}
    \right)\tf - \D^2\tf - 2\g\left({\bf M}\D + \frac{\nu}{1-2\nu}
    \text{tr}\{{\bf M}\} \D \right)\cdot\tu\right],\\
  &(1-2\nu)\D^2\tu + \D(\D\cdot\tu) = 2\left[(1-2\nu){\bf M}\D + \nu\, 
    \text{tr}\{{\bf M}\} \D\right]\tf.
\end{align}
By introducing the vector of fields $\tilde{\bf v} = (\tilde{u}_1, \tilde{u}_2,
\tilde{u}_3, \tf)$, we can again take the normal mode of the perturbations as 
$\tilde{\bf v} = \hat{\bf v}\exp(\sigma t + i{\bf q}\cdot\x)$ and obtain a linear
algebraic system of the form ${\bf \Lambda}\hat{\bf v} = {\bf 0}$. This operator 
can then be written as ${\bf\Lambda}(\s,{\bf q}) = {\bf\Sigma}(\s) + {\bf Q}
({\bf q})$, where $\Sigma_{44} = \s$ is the only nonzero element of 
${\bf\Sigma}$, and the elements of ${\bf Q}$ are given in the Appendix. For 
the perturbation eigenvector $\hat{\bf v}$ to have nontrivial solutions, the
operator ${\bf \Lambda}$ must be singular ({\it i.e.~}$\text{det}\{{\bf\Lambda}\}
 = 0$), and hence we obtain the dispersion 
relation for the growth rate of a given spatial mode
\begin{equation}
  \s({\bf q}) = \frac{-\text{det}\{{\bf Q}({\bf q})\}}
     {\text{det}\{{\bf Q}^{(4)}({\bf q})\}},
\end{equation}
where the parenthetical superscript denotes the submatrix of ${\bf Q}$ obtained 
by removing the fourth row and column. A typical plot of $\s$ can be seen in
Figure (\ref{disp5}). For presentation purposes, we have only shown a slice in the 
$(x,y)$ plane. Lighter colors denote larger growth rates, the curve represents
the zero (neutral) contour of $\s$, and the $\times$'s mark the points of greatest
growth. The parameters are the same as those used in Figure (\ref{spin}) with 
the addition of lattice mismatch rates listed in the caption obtained by 
\cite{chen06}.
Note that the most excited wave-vectors lie along the $x$-axis, which for the 
parameters used, is the direction of greatest strain. 
To further understand this, we take the simple case of unidirectional strain 
($M_{11} = \Mu$, $M_{22} = M_{33} = 0$). If we then write the growth rate as 
$\s({\bf q}) = \s_0({\bf q}) + \s_s({\bf q})$, where $\s_0$ is the growth 
in the absence of strain ({\it i.e.}~$\Mu \rightarrow 0$), and $\s_s$ is the strain 
induced contribution, then this contribution takes the form
\begin{equation}
  \s_s({\bf q}) = -\frac{2\g\f_0 \Mu^2}{(1-\nu)q^2}\left(q_y^2+q_z^2\right)^2,
  \quad q = |{\bf q}|.
\end{equation}
It can be seen here that this contribution is minimized for a given wave
vector of magnitude $q$ in the $(q_y,q_z)$ plane, which means that 
phase-separation is suppressed the most in directions orthogonal to the
unidirectional strain. Additionally, along the vector ${\bf q} = (q_x,0,0)$,
the strain induced contribution is maximized at $\s_s = 0$, which means 
that the direction of strain for this case is the only direction without 
suppression.  It can be shown that if all $M_{ii}$ have the same sign 
({\it i.e.}~pure expansions or pure contractions), then phase-separation
will always occur in the direction in which $|M_{ii}| > |M_{jj}|, |M_{kk}|$
({\it i.e.}~the direction of largest strain), however it will be seen in the next
section that the presence both contractions and expansions result in 
the possibility of skewed phase interfaces.
  
\section{Numerical Simulations}
To verify the results from the linear stability analysis, we perform 
numerical simulations of the governing equations (\ref{bulk1}-\ref{bulk2}),
as this analysis can only predict the initial dynamics still within
the linear regime.  Both two- and three-dimensional simulations are
presented to represent two limits of this system.

 \begin{figure}[h] \centering
  \resizebox{6in}{!}{\includegraphics[width=4.5in]{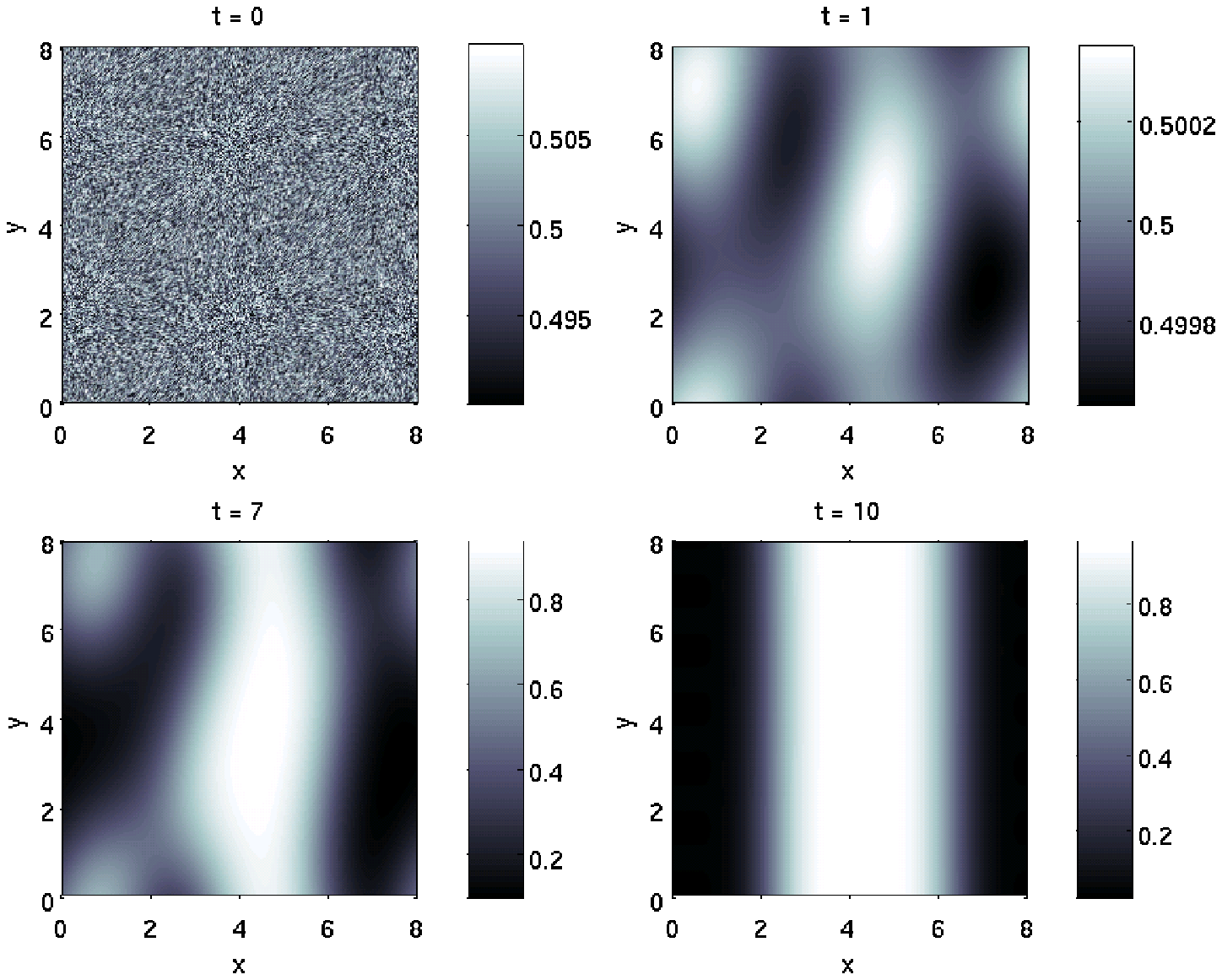}}
  \caption[Numerical simulation of system (\ref{bulk1}-\ref{bulk2})]
  {Time evolution of system (\ref{bulk1}-\ref{bulk2}) starting from 
    the state $\f = 0.5$ perturbed with small amplitude noise at 
    $t = 0,1,7,10$ as labeled.}
  \label{evo}
\end{figure}

\subsection{ Phase Separation with Anisotropic Plane Strain }
We first present the limit in which strains along the $c$-axis are neglected
({\it i.e.}~$u_3 \rightarrow 0$) under the assumption that Li transport along 
molecular channels direction is in constant equilibrium due to the very fast mobility 
\cite{amin07} and suppressed phase separation~\cite{Cogswell2012}. This is a reasonable assumption for reaction-limited \fepo nano particles~\cite{singh08,Bai2011}, but would break down in larger particles ($> 100$nm) due to channel-blocking Li/Fe anti-site defects, which lead to size-dependent effective mobility~\cite{Malik2010}. 
Note that the stability results from section (\ref{anisomm}) remain unchanged
for the two-dimensional case. For numerical convenience, we will further 
approximate the flux (before the rescaling) as ${\bf j} = -B\D\mu$. 
This is a common
assumption in phase-field modeling since the prefactor $\f$ merely scales the initial time evolution and 
does not affect long-term behavior of the system which is solely determined
by the form of the chemical potential \cite{han04, gurtin96, cahn61}. We use 
a pseudo-spectral method for the spatial differentiation with $256\times 256$ 
modes and a semi-implicit scheme for temporal
updating. The linear terms are handled with the backward Euler method, which 
although only first order accurate in time, provides the necessary stability 
required for the stiff solutions of the system. The second order Adams-Bashforth 
method is used for the nonlinearities, and periodic boundary conditions are used to 
approximate the infinite domain. As the 
equations for the deformation fields are linear, they can be solved exactly
in Fourier space in terms of the order parameter, and hence the system can be
reduced to a single evolution equation. If we define the Fourier transform of 
the order parameter as $\mathfrak{F}\{\f(\x,t)\}\equiv \phi({\bf q},t)$, where 
${\bf q}$ is the corresponding wave-vector, the evolution of $\phi$ in Fourier 
space will be therefore governed by
\begin{align}
  \frac{d\phi}{dt} = \;&\left(2\al q^2 - q^4 + \frac{\g\mathcal{M}({\bf q})}
       {(1-\nu)(1-2\nu)q^2}\right)\phi - q^2\mathfrak{F}\left\{\log
  \left(\frac{\f}{1-\f}\right)\right\},
\end{align}
where the contribution from the coherency strains is given as
\begin{align}
  \mathcal{M}({\bf q}) = \;&\left[(1-\nu)M_{11}+\nu M_{22}\right]\left[
    \left((1-\nu)q^2+q_y^2\right)M_{11} + \left(\nu q^2-q_y^2\right)
    M_{22}\right]q_x^2\\
  &+ \left[\nu M_{11}+(1-\nu)M_{22}\right]\left[\left(\nu q^2-q_x^2
    \right)M_{11} + \left((1-\nu)q^2+q_x^2\right)M_{22}\right]q_y^2.
\end{align}

A typical simulation evolved from the state $\f = 0.5$ perturbed with 
small amplitude noise is shown at various times in Figure (\ref{evo}). These 
perturbations are initially damped, and regions of localized concentrations 
begin to form. The regions coarsen until the concentration is aligned with
the $y$-axis confirming the spontaneous phase-separation along the direction 
of greatest strain.
Note that the parameters are the same as those listed in the captions 
of Figures (\ref{spin}) and (\ref{disp5}) excluding contributions from the third 
dimension. It should be mentioned that this model is quite robust, as similar 
results can be obtained from simulations over a wide range of parameters.

 \begin{figure}[h] \centering
   \resizebox{3in}{!}{\includegraphics[width=3in]{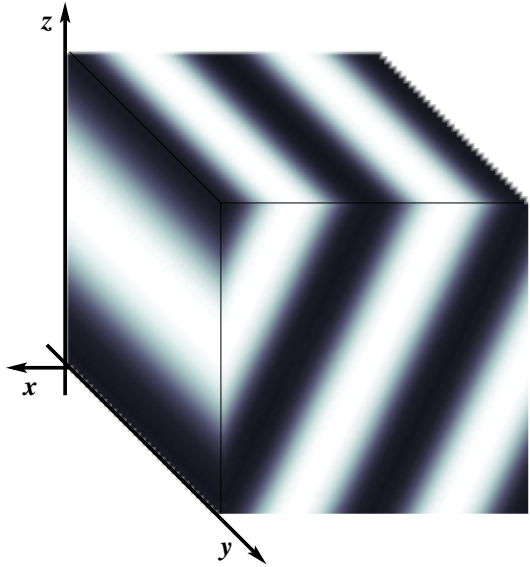}}
   \caption{Equilibrium state of \fepo with anisotropic coherency strain in three dimensions showing skewed, striped phases. }
   \label{cube}
\end{figure}

\subsection{ Three Dimensional Coherency Strain in \fepo}
For a more complete picture of phase separation in \fepo, it is necessary to consider fully anisotropic three-dimensional coherency  strain. In particular, we perform full three-dimensional simulations to include the effects
of contractions in the $c$-direction.  As seen in Figure (\ref{cube}), the 
equilibrium state results in skewed interfaces, which have also been seen in some experiments~\cite{laffont06,Ramana2009}.

\section{Conclusions}
We have considered the intercalation of Li within \fepo~as a model system 
which exhibits both phase-separation and highly anisotropic coherency
strains. A phase-field model which incorporates energetic contributions from 
entropy, enthalpy and elastic properties of the host material has been
developed.  We have shown through linear stability analysis that while coherency
strains suppress phase-separation, this suppression is maximized in directions 
orthogonal to the direction of a given strain. Hence the system is expected to
phase-separate along the direction of greatest strain ($a$-axis) given sufficiently  
large enthalpic contributions.  As linear stability theory can only predict the initial 
evolution of spinodal decomposition, numerical simulations were performed on an 
idealized system to illuminate the effects of coherency strain anisotropy, and phase 
boundaries were observed to align perpendicularly to the $a$-axis as observed 
in experiment. Coherency strain anisotropy has therefore been shown to be a possible 
mechanism for this alignment. We have also considered phase separation in the three dimensional case with fully anisotropic coherency strain, including contraction along the $c$ axis. This leads to skewed phase boundaries, which have also been seen in some experiments. These interesting observations are  unified and extended to electrochemically driven intercalation at fixed current in Ref.~\cite{Cogswell2012}.

\section*{Acknowledgements}
This work was supported by the National Science Foundation under Contracts DMS-0842504 and DMS-0948071 and partially under the auspices of the US Department of Energy by LLNL 
under Contract DE-AC52-07NA27344.

\section*{Appendix}
The elements of ${\bf Q}$ are listed as follows, where we have defined
$q \equiv |{\bf q}|$ with ${\bf q} = (q_x,q_y,q_z)$.
\begin{align*}
  &Q_{11} = (1-2\nu)q^2 + q_x^2,\quad Q_{12} = q_xq_y,\quad 
  Q_{13} = q_xq_z,\quad
  Q_{14} = 2i[(1-2\nu)M_{11} + \nu M_{kk}]q_x,\\
  &Q_{21} = q_yq_x,\quad Q_{22} = (1-2\nu)q^2 + q_y^2,\quad 
  Q_{23} = q_yq_z,\quad
  Q_{24} = 2i[(1-2\nu)M_{22} + \nu M_{kk}]q_y,\\
  &Q_{31} = q_zq_x,\quad Q_{32} = q_zq_y,\quad 
  Q_{33} = (1-2\nu)q^2 + q_z^2,\quad
  Q_{34} = 2i[(1-2\nu)M_{33} + \nu M_{kk}]q_z,\\
  &Q_{41} = -2i\g\f_0\left(M_{11} + \frac{\nu M_{kk}}{1-2\nu}
  \right)q^2q_x,\quad
  Q_{42} = -2i\g\f_0\left(M_{22} + \frac{\nu M_{kk}}{1-2\nu}
  \right)q^2q_y,\\
  &Q_{43} = -2i\g\f_0\left(M_{33} + \frac{\nu M_{kk}}{1-2\nu}
  \right)q^2q_z,\quad
  Q_{44} = -\f_0\left(2\al - \frac{1}{\f_0(1-\f_0)} - 
  q^2\right)q^2.
\end{align*}

\end{document}